\documentclass[,showpacs,amsmath,amssymb,prl]{revtex4}

\usepackage{graphicx}
\usepackage{amsmath}
\usepackage{amssymb}

\begin{document}

\title{Spectral flow in superconducting point contacts}
\author{N. B. Kopnin $^{(1,2,3)}$}
\author{V. M. Vinokur $^{(3)}$}
\affiliation{$^{(1)}$ Low Temperature Laboratory, Helsinki
University of
Technology, P.O. Box 2200, FIN-02015 HUT, Finland,\\
$^{(2)}$ L. D. Landau Institute for Theoretical Physics, 117940
Moscow, Russia\\
$^{(4)}$ Argonne National Laboratory, Argonne, Illinois 60439 }

\pacs{74.80.Fp, 73.23.-b, 73.63.Rt}

\begin{abstract}
We find that multiple Andreev reflections mediating the transport
in superconducting point contacts are strongly affected by a small
amount of impurities in the area of the contact. We also argue
that the model based on Zener transitions within independent
conducting channels is not suitable for kinetic processes in
multichannel contacts.
\end{abstract}

\maketitle

\section{Introduction}

Effect of impurities on transport in superconducting
nanostructures is one of the key issues in the physics of
mesoscopic superconductors, among which point contacts are the
simplest devices of interest. One can distinguish contacts with
only a few transverse eigenmodes (or conducting channels) relevant
for a given temperature (quantum contacts) and those where the
number of channels $N_{\rm ch}$ is relatively large (mesoscopic
contacts). In quantum contacts, the effect of impurities on both
static and transport properties can be satisfactorily described by
introducing certain transmission and reflection probability
$D+R=1$ for each conducting channel \cite{Beenakker/rev}.
Mesoscopic weak links with disorder are sometimes also treated
within this scheme. To this end, a quantum eigenstate problem for
a charge carrier is to be solved for the particular realization of
disorder, each state being considered as an individual channel
with certain transmission probability. The problem of a
multichannel weak link is then mapped into that for a set of
tunnel contacts having an effective distribution of transmission
probabilities. This scheme has been shown recently
\cite{Kopnin/pointcont} to be equivalent to the semi-classical
theory of superconductivity for calculating some {\it static}
properties of multichannel contacts.

The individual channels found by solving the quantum-mechanical
problem are independent in the sense that transitions between them
are absent under static conditions. One might now raise a question
how useful is the model of independent channels for {\it
transport} properties of multichannel contacts when time-dependent
perturbations induce transitions between the individual channels.
This is of a crucial importance for further theoretical studies of
transport in superconducting mesoscopic multichannel devices in
presence of disorder. We address this problem using a specific
example of a ballistic point contact between two clean
superconductors assuming that the thickness $d$ of the insulating
layer and the size $a$ of the orifice in it are much larger than
atomic dimensions but shorter than the coherence length $\xi $ and
the mean free path $\ell$, i.e., $a\sim d\ll \xi \ll \ell $. The
contact thus has a very large number of conducting channels,
$N_{\rm ch}\sim \left(p_F a\right)^2\gg 1$. The spectrum of the
midgap states localized within the coherence length $\xi $ near
the point contact was first found in \cite{Kulik}
\begin{equation}
\epsilon _{\pm }=\mp |\Delta |\cos (\phi /2)  \label{spectr/clean}
\end{equation}
for a phase difference $\phi$ between the two superconductors. The
spectrum is degenerate: all particles flying from one
superconductor into another have the same energy independently of
their momentum directions within the corresponding hemispheres.
Another example is a contact having a tunnel barrier with a
transmission probability $D$. The spectrum \cite{barrierspect}
\begin{equation}
\epsilon =\pm |\Delta| \sqrt{1-D\sin ^2(\phi /2)}
\label{specrt/tunnel}
\end{equation}
has a (mini)gap $\sqrt{R}|\Delta|$ near the phase difference $\phi
=\pi$, which separates the two branches of the spectrum. The
mid-gap states have a profound effect on dynamics of the contacts.
When the applied voltage is larger than the inelastic relaxation
rate, $eV\gg \tau _{\epsilon}^{-1}$, a nonlinear regime sets in
which originates from multiple Andreev reflections (MAR)
\cite{Zaikin,Cuevas,AverinBardas2,AverinBardas1,AverinBardas3,Bratus}
of localized quasiparticles. For a contact without both impurities
and a barrier, excitations from below the gap edge with $\epsilon
=-|\Delta|$ are accelerated through the midgap states Eq.\
(\ref{spectr/clean}) and re-appear above the gap with $\epsilon
=|\Delta|$ and vice versa, thus producing a highly nonequilibrium
situation. On the contrary, if a contact has a barrier with a
reflectivity $R$, the localized excitations cannot accelerate due
to the minigap and retain their original states. It is only Zener
tunnelling through the minigap that provides nonequilibrium
current through the contact
\cite{AverinBardas1,AverinBardas3,Bratus}.

The model of independent channels was applied to kinetic effects
in multichannel point contacts in Ref.\ \cite{AverinBardas3}. The
key ingredients of that model were (i) the MAR processes in
separate channels with (ii) Zener transitions through a minigap
between two spectral branches in Eq.\ (\ref{specrt/tunnel}) within
each individual channel, and (iii) a standard distribution
\cite{Dorokhov} of reflection coefficients over the channels. Here
we study the transport in mesoscopic multichannel contacts with
impurities and show that the above model of individual channels
consisting of processes (i)--(iii) is not suitable for dynamic
phenomena when the number of channels is large, $N_{\rm ch}\gg 1$,
because transitions between different channels are excited in
time-dependent conditions. As a result, the system finds the
optimal way to relax through channels with small reflection
probabilities avoiding weak Zener tunneling. We further
demonstrate that the concept of MAR is equivalent to the {\it
spectral flow} appearing in a wide variety of quantum systems
\cite{specflow,KopVol/specflow,Stone}. Using the quasiclassical
theory of nonstationary superconductivity we show that the
impurity scattering provides an effective momentum exchange
between particles moving from the opposite sides of the contact
and impedes the MAR acceleration. For low inelastic relaxation
rate, the I--V curve displays a non-trivial behavior: At low
voltages, impurity relaxation slows the acceleration down and
results in a new linear current-voltage dependence with a large
conductance. This differs from a square-root dependence obtained
within the independent-channel model (i)--(iii).

\section{Kinetic equation}

The density of states (DOS) for the midgap states in a classical
point contact was calculated in Refs.\
\cite{Kulik,Kopnin/pointcont}.  It is $\nu (s)=\nu (0)e^{-k|s|}$
with $k =2v_{F}^{-1}\sqrt{|\Delta |^{2}-\epsilon ^{2}}$ where $s$
is the distance along the particle trajectory; $\nu $ is
normalized to the normal single-spin DOS at the Fermi level,
$N_F$. We use the standard kinetic equation approach \cite{LO,KL}
where the distribution function is split into two components
$f^{\left(1\right)}$ and $f^{\left( 2\right) }$ that are
respectively odd and even in the variables $\epsilon,\, {\bf p} $.
They are coupled by the Boltzmann equation
\begin{equation}
\nu v_F \frac{1}{2} \frac{\partial \dot{\chi}}{%
\partial s}\frac{\partial  f^{(1)}}{\partial \epsilon }
+\nu \frac{\partial  f^{(1)}}{\partial t}+v_{F}%
\frac{\partial }{\partial s}\left( \nu  f^{(2)}\right)
= J~.  \label{kineq3}
\end{equation}
The second equation ensures that the distribution function $
f^{\left( 1\right) }$ is constant along the trajectory \cite{KL}.
In Eq.\ (\ref{kineq3}), the first term in the l.h.s. originates
from the Doppler energy and $\dot{\chi}$ is the time derivative of
the phase, and $J$ is the impurity collision integral. A detailed
discussion of the kinetic equations is given in Refs.\
\cite{LO,KL}. The standard technique of averaging over impurities
applies since the number of impurities within the volume of the
orifice is large. Indeed, the mean free time is $\tau ^{-1}\sim
N_{F}n_{{\rm imp}}|u|^{2}$ where $|u|\sim Up_{F}^{-3} $ is the
Fourier transform of the impurity potential $U$. The number of
impurities $n_{{\rm imp}}da^{2}\sim (d/\ell )(E_{F}/U)^{2}N_{\rm
ch}$ can be very large even for $U\sim E_{F}$ because of a
macroscopic number of quantum channels in the orifice $N_{\rm ch}
\gg 1$. The large number of impurities is the essential parameter
in our theory.

We integrate Eq.\ (\ref{kineq3}) along the trajectory across the
contact over distances longer than $\xi $. The first term in the
l.h.s. of Eq.\ (\ref{kineq3}) has a sharp maximum at $s\sim d$
where $\nu (s)$ is nearly constant. The third term in the l.h.s.
disappears due to the decay of $\nu $. The collision integral
contains \cite{Kopnin/pointcont} a combination of angular averages
$\nu \left[ \left\langle \nu \right\rangle %
f^{(1)}-\left\langle \nu  f^{(1)}\right\rangle \right] $. We write
 $\left\langle \nu \right\rangle =\left[
\left\langle \nu \right\rangle _{+}+\left\langle \nu \right\rangle
_{-}\right] /2$ where $
\left\langle \nu \right\rangle _{\pm }=(2\pi )^{-1}\int_{\pm p_{x}>0}\nu (%
{\bf p},{\bf r})\,d\Omega _{{\bf p}}
$. The upper (lower) sign is for right-moving (left-moving) particles
with $p_x>0$ ($ p_x<0$). We choose the direction of the $x$ axis
from the region with the phase $\chi_1$ into that with $\chi_2$.
Since $ f^{(1)}$ depends on the sign of $%
p_{x}$ only, being independent of the direction of ${\bf p}$
within each hemisphere $p_x >0$ or $p_x<0$,
\begin{equation}
J_{\pm }=-\tau ^{-1}\nu _{\pm }\left\langle
\nu \right\rangle _{\mp }\left(  f_{\pm }^{(1)}- f_{\mp
}^{(1)}\right) .\label{collintegral}
\end{equation}
The collision integral Eq.\ (\ref{collintegral}) is only
nonvanishing if $\nu _+$ and $\nu _-$
are nonzero simultaneously. It is localized at distances $s\sim d$:
The averages $%
\left\langle \nu \right\rangle $ decay at $s\gtrsim a$ away from
the contact, being proportional to the solid angle at which the
orifice is visible from the position point. Since $\nu$ is
constant within the orifice, $ \ell ^{-1}\int ds\left\langle \nu
\right\rangle _{\mp }=\gamma \nu _{\mp }(0) $ where $\gamma \sim
d/\ell $ is a contact-dependent geometric factor; we assume a
constant $\gamma $ for simplicity. Finally, putting $\phi =\chi
_{2}-\chi _{1}$ we find
\begin{equation}
\nu _\pm (0)\left[ \pm \frac{1}{2}\frac{\partial \phi }{\partial
t}\frac{\partial
 f_\pm ^{(1)}}{\partial \epsilon }+\frac{1}{\sqrt{|\Delta |^2-
 \epsilon ^2}} \frac{\partial  f_\pm %
^{(1)}}{\partial t}\right]
 =-\gamma \nu _{\pm }(0)\nu
_{\mp }(0)\left( f_{\pm }^{(1)}- f_{\mp }^{(1)}\right) .
\label{kineq/integr1}
\end{equation}

\section{Spectral flow in a ballistic contact}

In the collisionless limit \cite{Kulik},
\begin{equation}
\nu _\pm \left( 0\right) =\pi \sqrt{|\Delta |^{2}-\epsilon
^{2}}\delta \left( \epsilon -\epsilon _{\pm
}\right) . \label{dos}
\end{equation}
Since $\partial \epsilon _{\pm }/\partial \phi =\pm
\frac{1}{2}|\Delta |\sin (\phi /2)$ we can write Eq.
(\ref{kineq/integr1}) as $\nu _\pm (0) df^{(1)}/dt =0 $ where the
total time derivative of the distribution function is taken along
the spectrum Eq. (\ref{spectr/clean}). The total time derivative
vanishes: the distribution is constant in time and thus it is
independent of energy. To find it we note that, as $\phi $ varies
from $0$ to $2\pi $, the distribution of particles with $p_{x}>0$
remains the same as it was at $t=0,\;\phi =0$ for an energy
$\epsilon =-\left| \Delta \right| $, i.e., $ f^{(1)}=- f_{\Delta
}$ where $ f_{\Delta }=\tanh \left( |\Delta |/2T\right) $.
For particles with $p_{x}<0$, the distribution remains as it was at $%
t=0,\;\phi =0$, and $\epsilon =\left| \Delta \right| $, i.e., $ f^{(1)}=%
 f_{\Delta }$. If $\phi $ decreases from $2\pi $ to $0$, the signs are
opposite,
\begin{equation}
 f_{\pm }^{(1)}=\mp  f_{\Delta }\,{\rm sign}\,(\dot{\phi}) .
\label{f/ballistic}
\end{equation}
Excitations are accelerated by the applied voltage undergoing
multiple Andreev reflections \cite{Zaikin,Cuevas,AverinBardas1}
before their energies grow by $2|\Delta |$ to let them escape from
the potential well.

\begin{figure}[t]
 \centerline{\includegraphics[width=0.3\linewidth]{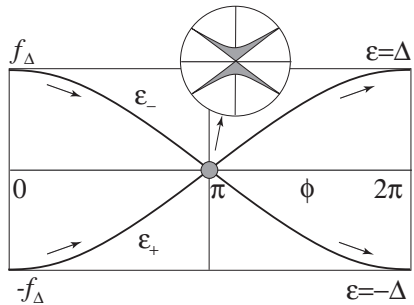}}
 \caption{Spectral flow through the midgap states. For $\dot{\phi}>0$, a
 particle with $p_x>0$ is captured at $\phi =0$ with an energy
 $\epsilon _+ =-|\Delta|$ and then released at $\phi =2\pi$ with
 $\epsilon _+ =|\Delta|$; a particle having $p_x<0$ is captured with
 $\epsilon _- =|\Delta|$ and released with $\epsilon _-
 =-|\Delta|$. The dark dot (expanded in the insert)
 shows the impurity band $\sim |\Delta|\sqrt{\gamma}$ near
 the crossing point.}
 \label{scsflow}
 \end{figure}

The same process can also be viewed differently. Under a small
voltage bias $eV\ll |\Delta |$, quasiparticles are accelerated
along the spectrum Eq.\ (\ref{spectr/clean}) with $\phi =2eVt$. As
the phase varies from $0$ to $2\pi $, the quasiparticle energy
changes from $\mp |\Delta |$ to $\pm |\Delta |$. A quasiparticle
that has been captured below the gap edge emerges above the gap
after one period of Josephson oscillations (see Fig.\
\ref{scsflow}). Another particle gets captured and then released
during the next period, and so on \cite{AverinBardas2}. This is
known as the {\it spectral flow}. It plays a crucial part in
vortex dynamics \cite{KopVol/specflow,Stone}. In a context of weak
links, it was first discussed in Ref.\ \cite{Volovik}. In the
absence of collisions, the distribution of excitations localized
near the contact Eq.\ (\ref{f/ballistic}) is determined by
distribution of excitations at the gap edges, which are
delocalized and are thus nearly in equilibrium due to the
relaxation in the bulk. However, emerging quasiparticles have a
``wrong'' distribution with respect to the ``native'' ones; they
relax producing dissipation.


The current through the contact with the Sharvin conductance
$R_{N}^{-1}=e^{2}N_{F}v_{F}S/2$ is
\begin{equation}
I=-\frac{1}{eR_N}\int \frac{d\epsilon}{2}\left[ \nu_+  f_+^{(1)}
-\nu _-  f_-^{(1)} \right] . \label{Igeneral}
\end{equation}
where $S\sim a^2$ is the area of the contact. With Eq.\
(\ref{f/ballistic}) the current due to Andreev states is
\begin{equation}
|e|R_N I=\pi |\Delta |\left|\sin (\phi /2)\right|
 f_\Delta 
\, {\rm sign}(V) . \label{I/ballistic}
\end{equation}
Excitations with energies $|\epsilon |>|\Delta |$ give a much
smaller current, which can be  neglected. Equation
(\ref{I/ballistic}) agrees with the results of Refs.\
\cite{Zaikin,Cuevas,AverinBardas1}.

\section{ Spectral flow and impurities}

A small amount of impurities in the area of the contact lifts the
degeneracy of Eq.\ (\ref{spectr/clean}) and modifies the spectrum
in a way that an impurity band appears near the point $\phi =\pi $
where the two spectral branches cross \cite{Kopnin/pointcont}. In
the presence of impurities the DOS is
$\nu _{\epsilon \pm }(0)=\left(E_{\mp }/2\gamma \right)%
\sqrt{\left(4\gamma / E_{+}E_{-}\right)-1} $ where $E_\pm
=\epsilon /|\Delta | \pm \cos (\phi /2)$. Both $\nu _+ $ and $\nu
_-$ are nonzero near the lines $E_\pm =0$ within the impurity band
$0<E_{+}E_{-}<4\gamma $ shown in Fig.\ \ref{scsflow}.
To simplify the partial differential equation
(\ref{kineq/integr1}) we note that, in the region far from the
crossing point, $|\epsilon |\gg |\Delta |\sqrt{\gamma}$, $|\phi
-\pi |\gg\sqrt{\gamma}$, the ratio $\nu _-/\nu _+=E_+/E_-$ is of
the order of $\gamma |\Delta |^2/ \epsilon ^2 \ll 1$ near the line
$E_+=0$, and it is $\nu _-/\nu _+ \sim \epsilon ^2/\gamma |\Delta
|^2\gg 1$ near $E_-=0$. Therefore, far from the crossing point,
$\left| \phi - \pi \right| \gtrsim \beta$ where $\sqrt{\gamma} \ll
\beta \ll \pi$, the DOS has the form of Eq.\ (\ref{dos}), and the
l.h.s. of Eq.\ (\ref{kineq/integr1}) reduces to the full
derivative $d f_\pm ^{(1)}/dt$ along the respective lines $E_\pm
=0$. Since the collision integral also vanishes for $\left| \phi -
\pi \right| \gtrsim \beta$, the distribution is constant in time
far from the crossing point, in the same way as it was without
impurities.

Equation (\ref{kineq/integr1}) can be solved by perturbations in
the limit of small voltages. The zero-order term corresponds to
vanishing of the l.h.s.\ and results in the condition $f_{+
}^{(1)} = f_{-}^{(1)}$ that holds through the crossing point for a
given energy. We find that the distribution functions at the four
lines in Fig.\ \ref{scsflow} entering the crossing region satisfy
$f^{(1)}_{\pm} (\pi +\beta)=f^{(1)}_{\mp} (\pi -\beta)=\pm
f_{\Delta}$. To the zero approximation, particles emerging in the
continuum with an energy $\epsilon =\pm |\Delta |$ have the
distribution $\pm f_{\Delta}$ that corresponds to the local
equilibrium at the gap edges
\begin{equation}
 f_{\pm }^{({\rm eq})}=\mp {\rm sign}[\cos (\phi /2)]\,\, f%
_{\Delta }.  \label{equildist}
\end{equation}
In this approximation, no relaxation in the continuum occurs. The
deviation $f_{1\pm}$ from the local equilibrium is proportional to
the time derivative, i.e., to the voltage. We write $f_{\pm}^{(1)}
= f^{({\rm eq})}_\pm +f_{1\pm}$. The l.h.s. of Eq.\
(\ref{kineq/integr1}) can be estimated as $\nu \, eV
\left(f_{\Delta}/\epsilon ^*\right)$ where $\epsilon ^* \sim
|\Delta |\sqrt{\gamma}$ is the characteristic scale of energy near
the crossing point. Using $\nu \sim 1/\sqrt{\gamma}$ we find
$f_{1\pm} \sim \mp \left( eV/|\Delta |\gamma \right)f_{\Delta}$.
The deviation $f_{1\pm}$ is thus linear in $V$ for low voltages.

In general, $ f_\pm ^{(1)}$ can be found by matching the four
constants $f^{(1)}_{\mp} (\pi \pm\beta)$ while integrating Eq.
(\ref{kineq/integr1}) across $\phi =\pi$. We note that for a
contact with a barrier of reflectivity $R=1-D$, the excitation
spectrum is given by Eq.\ (\ref{specrt/tunnel}) with a gap
$\sqrt{R}|\Delta|$ that separates states with positive energies
from those with negative energies. As a result, the distribution
for excitations with $\epsilon >0$ is decoupled from that for
$\epsilon <0$ unless the Zener tunneling mixes them
\cite{AverinBardas1,AverinBardas3}. On the contrary, in
macroscopic contacts with impurities, the states near $\phi =\pi$
form a continuum where the regions $\epsilon >0$ and $\epsilon <0$
are connected along the lines of a nonzero DOS passing through
$\phi =\pi$, see Fig.\ \ref{scsflow}. According to Eq.\
(\ref{kineq/integr1}), the distribution of excitations is thus
continuous within the entire region of nonzero DOS. Therefore, the
distributions on all four lines entering the crossing region in
Fig.\ \ref{scsflow} are coupled to each other.

Since we are interested the asymptotic values of $ f_\pm ^{(1)}$
for  $\phi =\pi -\beta $ and $\phi =\pi +\beta $, we neglect the
fine structure of $\nu $ and approximate it by Eq.\ (\ref{dos}).
The collision integral is then nonzero only at the crossing point
of $\epsilon _+$ and $\epsilon _-$ defined by Eq.\
(\ref{spectr/clean}). Equation (\ref{kineq/integr1}) yields
\begin{equation}
\frac{d f_{\pm }^{(1)}}{dt}=\mp \pi\gamma |\Delta|^2
\Gamma \delta\left(\epsilon -\epsilon _\mp \right) .
\label{kineq/scs/integrated}
\end{equation}
$\Gamma$ is a combination $f^{(1)}_+(\epsilon ,\phi
)-f^{(1)}_-(\epsilon ,\phi )$ taken in the vicinity of $\epsilon
=0$, $\phi =\pi$, which couples the four values $f^{(1)}_\pm (\pi
\pm \beta )$. We shall see that $\Gamma$ varies slowly near $\phi
=\pi$ thus we approximate
\[
\Gamma =\frac{1}{2}%
\left[ f_{+}^{(1)}(\pi +\beta )+ f_{+}^{(1)}(\pi -\beta ) -
f_{-}^{(1)}(\pi +\beta )- f_{-}^{(1)}(\pi -\beta )\right] .
\]
We integrate Eq.\ (\ref{kineq/scs/integrated}) over time along
the dependence $E_+=0$ for particles with $p_{x}>0$ or along $E_-=0$ for
$p_{x}<0$ and obtain
\begin{equation}
 f_{\pm }^{(1)}(\pi +\beta )\!- f_{\pm
}^{(1)}(\pi -\beta ) =\mp \alpha \Gamma
\label{kineq/scs/integrated1}
\end{equation}
where $\alpha =\pi |\Delta |\gamma /2eV$. Adding the two Eqs.\
(\ref {kineq/scs/integrated1}) we find that the quantity
\begin{equation}
A(\beta )=f_{+}^{(1)}(\pi +\beta )- f_{-}^{(1)}(\pi -\beta)
\label{combin1}
\end{equation}
satisfies $A(\beta )=A(-\beta)\equiv A$, i.e., it is a slowly
varying function near $\beta =0$. This proves that $\Gamma
=[A(\beta )+A(-\beta)]/2$ can indeed be treated as a constant,
$\Gamma =A$.

Consider first the case $\dot{\phi}=2eV>0$. The initial condition at $t=0$, $%
\phi =0$ is the equilibrium distribution at an energy $\epsilon =\mp |\Delta
|$, i.e., $ f_{\pm }^{(1)}=\mp  f_{\Delta }$. Since the
distribution is constant outside the crossing region, we have also $ f%
_{\pm }^{(1)}(\pi -\beta )=\mp f_{\Delta }$. To find the functions $ f%
_{\pm }^{(1)}(\pi +\beta )$ we use Eqs.\ (\ref{kineq/scs/integrated1},
\ref{combin1}) whence $A=-2  f_{\Delta }/\left( 1+\alpha
\right)$ and
\[
 f_{\pm }^{(1)}(\pi +\beta )=\mp \left[\left(1-\alpha
\right)/\left(1+\alpha \right)\right] f_{\Delta } .
\]
For weak scattering $\alpha \ll 1$, the distribution remains
unchanged as compared to its initial value, $ f_{\pm }^{(1)}(\pi +\beta )= f%
_{\pm }^{(1)}(\pi -\beta )=\mp  f_{\Delta }$. Quasiparticles relax
towards the local equilibrium $\pm f_\Delta$
as they emerge in the continuum from inside the gap at an energy $%
\pm |\Delta |$. If scattering is strong $\alpha \gg 1$, the
distribution changes considerably at the crossing point $\phi =\pi
$ due to exchange between particles moving from the opposite sides
of the contact. Emerging particles with an energy $\epsilon =\pm
|\Delta |$ have a distribution close to the ``local equilibrium''
Eq.\ (\ref{equildist}) such that no relaxation occurs. This is not
the true equilibrium for $|\epsilon |<|\Delta|$: the latter can
only be established when the spectral flow rate $eV$ is much
smaller than the inelastic rate. The local equilibrium is achieved
through a very effective branch mixing due to the momentum
exchange on impurities when particles pass through the crossing
region of the spectral branches. The deviation from the local
equilibrium is zero as long as $\phi <\pi $ while it appears for
$\phi >\pi $:
\begin{equation}
 f_{1\pm }\equiv  f_{\pm }^{(1)}(\pi +\beta )-\left[ \pm  f%
_{\Delta }\right] =\mp \left[2/\left(1+\alpha \right)\right]
f_{\Delta } . \label{deviation}
\end{equation}
This agrees with our earlier perturbative estimate for the
low-voltage limit.

Calculating the current for $\phi <\pi $ from Eq.\ (\ref{Igeneral}) we find
$I=I_{s}$ where the ``supercurrent'' is defined through the
``equilibrium distribution'' Eq.\ (\ref{equildist}):
\[
eR_NI_{s}=\pi |\Delta |\sin (\phi /2)f_\Delta \,
{\rm sign}\left[ \cos (\phi /2)\right] .
\]
This expression differs from the well-known result for a
steady-state supercurrent \cite{Kulik}. The reason is that Eq.\
(\ref{equildist}) is not the true equilibrium, see the discussion
above. For $\phi >\pi $ we have $I=I_{s}+I_{n}$; the ``normal''
current contains the deviation from equilibrium Eq.\
(\ref{deviation}). Considering also the case $\dot{\phi}=2eV<0$ we
find for the normal current
\[
|e|R_N I_{n}=2\pi |\Delta |\left(1+|\alpha |\right)^{-1}\left| \sin (\phi
/2)\right|f_\Delta \,  {\rm sign}(V) .
\]
As in Ref.\ \cite{AverinBardas3}, it exists only
when $\pi <\phi <2\pi $ for $\dot{\phi%
}>0$ or when $0<\phi <\pi $ for $\dot{\phi}<0$.

For a voltage-biased contact, the supercurrent vanishes after
averaging over time:
\begin{equation}
|e|R_N \overline{I}=2|\Delta |\left(1+|\alpha |\right)^{-1} f_\Delta
\, {\rm sign}(V) . \label{normcurrent}
\end{equation}
For high voltages, $|\alpha |\sim|\Delta |\gamma /|eV|\ll 1$, the
MAR result Eq.\ (\ref{I/ballistic}) is reproduced. For low
voltages $|\Delta |\gamma \gg |eV|\gg \tau _{\epsilon}^{-1}$, the
current is linear, $\overline{I}=GV$ where $ G=(4/\pi \gamma
R_{N})f_\Delta $. This conductance is much larger than the
contribution $\sim R_N^{-1}$ from the states with $|\epsilon
|>|\Delta|$. To our knowledge, this low-voltage region has not
been studied experimentally though the contacts with suitable
parameters are now under intensive investigation \cite{contacts}.

\section{Discussion}

The linearity of the I-V curve for low voltages disagrees with a
square-root behavior predicted in \cite{AverinBardas3} using the
model based on Zener transitions within independent channels
averaged over the universal distribution of reflection
coefficients \cite{Dorokhov} valid for diffusive conductors. This
distribution cannot be applied directly for nearly ballistic
contacts. The proper distribution for contacts with a small
disorder was found \cite{Kopnin/pointcont} to be $P(R)= (2\pi
\gamma )^{-1}\sqrt{\gamma -R}/\sqrt{R}$. Its denominator does also
contain $\sqrt{R}$ like the universal distribution, which, at the
first glance, should have also resulted in a square-root
current-voltage dependence. However, the kinetic-equation approach
applicable for a multichannel contact leads to a different result.
The reason is that, in a macroscopic contact, the quantum channels
are very close to each other in energies with the level spacings
inversely proportional to the number of channels $N_{\rm ch}$.
Thus, in addition to the transitions within each channel (which
are the only ones considered in the independent-channel model),
time dependent processes excite even more effectively transitions
{\it between different channels}. As a result, the distribution
function is spread over the entire energy interval of nonzero DOS
or over the range $R\sim \gamma$ of reflection coefficients. The
role of each particular channel within the range $\delta R$ should
thus be weighted with $\delta R /\gamma$. If we now average the
low-voltage Zener processes having the rate $\propto \exp
(-|\Delta |R/eV)$ with $P(R )$ we would obtain the essential
reflection coefficients $R\sim eV/|\Delta|$. Thus the fraction of
essential channels contributing to the current in multichannel
contacts is decreased by a factor $ eV/|\Delta|\gamma$ as compared
to the independent channel model. This yields a much smaller
current compared to what is predicted by the kinetic-equation
approach. Therefore, the Zener processes can be neglected in favor
of the momentum relaxation on impurities. To summarize, the model
based on Zener transitions within independent channels does not
seem to be suitable for kinetic processes in multichannel
contacts. We believe, however, that calculations using the
Landauer approach with the full account of interchannel
transitions followed by the appropriate averaging would be
equivalent to the kinetic equation technique.

We thank D.\ Averin, M.\ Feigel'man, A.\ Larkin, and G.\ Volovik
for valuable discussions. This work was supported by the U.S.
Department of Energy Office of Science through contract No.
W-31-109-ENG-38 and by the Russian Foundation for Basic Research.


\begin{thebibliography}{99}

\bibitem{Beenakker/rev}  Beenakker C. W. J.,
Rev. Mod. Phys.{\bf 69}, 731, (1997).

\bibitem{Kopnin/pointcont} Kopnin N. B.,
Phys. Rev. B {\bf 65}, 132503, (2002).

\bibitem{Kulik} Kulik I. O. and Omel'yanchuk A. N.,
Fiz. Nizk. Temp.{\bf 3}, 945 (1977) [Sov. J. Low Temp. Phys. {\bf
3}, 459 (1977)].

\bibitem{barrierspect} Beenakker C. W. J., Phys.\
Rev.\ Lett. {\bf 67}, 3836, (1991).


\bibitem{Zaikin} Gunsenheimer U. and Zaikin A. D.,
Phys. Rev. B {\bf 50}, 6317, (1994).

\bibitem{Cuevas}  Cuevas J. C., Mart{\'{\i }}n-Rodero A., and
Levy Yeyati A., Phys. Rev. B {\bf 54}, 7366, (1996).

\bibitem{AverinBardas2}  Averin D. and Bardas A.,
Phys. Rev. B {\bf 53}, R1705 (1996).


\bibitem{AverinBardas1} Averin D. and Bardas A.,
Phys. Rev. Lett. {\bf 75}, 1831 (1995).

\bibitem{AverinBardas3} Bardas A. and Averin D.,
Phys. Rev. B {\bf 56}, R8518 (1997).

\bibitem{Bratus} Bratus' E. N., Shumeiko V. S., Bezuglyi E. V.
and Wendin G.,  Phys. Rev. B {\bf 55}, 12~666 (1997).


\bibitem{Dorokhov} Dorokhov O. N.,
Solid State Commun. {\bf 51}, 381 (1984).

\bibitem{specflow}  Callan C. G. and Harvey J. A.,
Nucl. Phys. B {\bf 250}, 427 (1985); Volovik G. E., Pis'ma Zh.
Eksp. Teor. Fiz. {\bf 43}, 428 (1986) [JETP Lett. {\bf 43}, 551
(1986)]; Stone M. and Gaitan F., Ann. Phys. (N.Y.) {\bf 178}, 89
(1987).

\bibitem{KopVol/specflow}  Kopnin N. B., Volovik G. E.,
and  Parts {\"{U}}. Europhys. Lett. {\bf 32}, 651 (1995).

\bibitem{Stone}  Stone M., Phys. Rev. B {\bf 54}, 13~222 (1996).


\bibitem{LO}  Larkin A. I. and Ovchinnikov Yu. N., in:
{\it Nonequilibrium Superconductivity} edited by D.\ N.\
Langenberg and A.\ I.\ Larkin,  (Elsevier Science Publishers,
Amsterdam, 1986) p. 493.


\bibitem{KL}  Kopnin N. B. and Lopatin A. V.,
Phys. Rev. B {\bf 51}, 15~291 (1995).

\bibitem{Volovik} Makhlin Yu. G. and Volovik G. E.,
Pis'ma Zh. Eksp. Teor. Fiz. {\bf 62}, 923 (1995) [JETP Lett. {\bf
62}, 941, (1995)].

\bibitem{contacts} Mur L. C., Harmans C. J. P. M., Mooij J. E.,
Carlin J. F., Rudra A., and Ilegems M., Phys. Rev. B {\bf 54}
R2327 (1996); Scheer E., Joyez P., Esteve D., Urbina C., and
Devoret M. H., Phys. Rev. Lett. {\bf 78}, 3535 (1997).


\end{thebibliography}
\end{document}